\documentclass[preprintnumbers, twocolumn,amsmath, amssymb, superscriptaddress, pra]{revtex4}
\usepackage{graphicx}
\usepackage{subfigure}
\usepackage{dcolumn}
\usepackage{bm}
\usepackage{epsf}
\usepackage{amssymb}
\DeclareGraphicsExtensions{.pdf}
\begin{document}

%

\author{David S. Simon}
\email[e-mail: ]{simond@bu.edu} \affiliation{Dept. of Physics and Astronomy, Stonehill College, 320 Washington Street, Easton, MA 02357} \affiliation{Dept. of
Electrical and Computer Engineering \& Photonics Center, Boston University, 8 Saint Mary's St., Boston, MA 02215, USA}
\author{Christopher R. Schwarze}
\email[e-mail: ]{crs2@bu.edu}\affiliation{Dept. of Electrical and Computer Engineering \& Photonics Center, Boston University, 8 Saint Mary's St., Boston,
MA 02215, USA}
\author{Alexander V. Sergienko}
\email[e-mail: ]{alexserg@bu.edu} \affiliation{Dept. of Electrical and Computer Engineering \& Photonics Center, Boston University, 8 Saint Mary's St., Boston,
MA 02215, USA} \affiliation{Dept. of Physics, Boston University, 590 Commonwealth Ave., Boston, MA 02215, USA}

\begin{abstract}
Grover multiports are higher-dimensional generalizations of beam-splitters, in which input to any one of the four ports has equal probability of exiting at any of the same four ports, including the input port. In this paper, we demonstrate that interferometers built from such multiports have novel features. For example, when combined with two-photon input and coincidence measurements, it is shown that such interferometers have capabilities beyond those of standard beam splitter-based interferometers, such as easily-controlled interpolation between Hong-Ou-Mandel (HOM) and anti-HOM behavior. Further, it is shown that the Grover-based analog of the Mach-Zehnder interferometer can make three separate phase measurements simultaneously. By arranging the transmission lines between the two multiports to lie in different planes, the same interferometer acts as a higher-dimensional Sagnac interferometer, allowing rotation rates about three different axes to be measured with a single device.
\end{abstract}

\title{Interferometry and higher-dimensional phase measurements with directionally-unbiased linear optics}


\maketitle
%

\section{Introduction}

 One of the chief characteristics of the Maxwell wave equation is that it is \emph{linear}, and that its solutions therefore obey the superposition principle: given two solutions to the wave equation, any linear combination of those solutions is also a valid solution. The result is that electromagnetic waves can interfere. More explicitly, since the intensity is proportional to the magnitude squared of the electric field, $I\sim |E|^2$, the intensity produced when two fields $E_1$ and $E_2$ are combined on a screen or viewing device is
\begin{equation}I\sim |E_1+E_2|^2 =I_1+I_2+2\; \mbox{Re}(E_1^\ast E_2).\end{equation} If the two fields differ in phase by $\phi$, the last term is the interference term, and is of the form \begin{equation}2|E_1E_2|\cos\phi \sim 2\sqrt{I_1I_2}\cos\phi .\end{equation} This classical interference has long formed the basis of a diverse array of precision measurement methods, with applications ranging from the determination of the structure of DNA via Bragg diffraction to the measurement of stellar radii with Michelson stellar interferometers.

Quantum mechanics is also based on a linear wave equation, the Schr\"odinger equation, so all quantum systems obey the superposition principle and exhibit interference between quantum amplitudes. As a result, optical interference effects can occur at the single photon level. So, for example, in a Young two-slit experiment a single photon impinging on a pair of slits has an amplitude for passing through each slit. As long as there is no measurement of which slit the photon passes through, the two amplitudes interfere with each other; after many trials, a visible interference pattern can be built up one photon at a time. Any measurement of which slit the photon traversed will destroy the interference pattern by effectively erasing one of the two interfering amplitudes. But as long as this ``which path'' information is absent, the resulting quantum interference pattern will look exactly like the classical interference pattern. This is \emph{single}-photon interference in the sense that what is interfering are multiple different amplitudes for the \emph{same} photon, not amplitudes for multiple different photons.

It is well-known that single-photon interference patterns can always be mimicked by interference of classical electromagnetic waves. To produce interference effects that are distinctly quantum in nature, it is necessary to go beyond single-photon interference. For two photon interference to occur, the state must be a superposition of two different two-photon states. Consider a \emph{pair} of photons (labeled by subscripts 1 and 2), that can be in two possible states (for example two polarization states or two spatial modes, labeled by the letters a and b). If there is no knowledge available about which photon is in which state, then the system is in a superposition of two possibilities:
\begin{equation}|\psi\rangle ={1\over \sqrt{2}} \left( |\psi_a\rangle_1|\psi_b\rangle_2 +e^{i\phi} |\psi_b\rangle_1|\psi_a\rangle_2\right) \label{mz}.\end{equation}
Such as state can be created, for example, using type II spontaneous parametric down conversion (SPDC) if the two oppositely-polarized photons are fed into separate arms of a Mach-Zehnder interferometer. If a polarization-dependent phase shift is included in the upper arm of the interferometer and we don't know which polarization traveled which path through the system, then the output state will be of the form above. The phase $\phi$ in equation \ref{mz} would then be the extra phase shift one polarization gains relative to the other when passing through the upper arm.
This two-particle superposition, which cannot be factored into a pair of single-particle states, is an example of an entangled state.
Interference between the pair of superposed two-photon amplitudes leads to effects that cannot be mimicked classically.

\begin{figure}
\centering
\includegraphics[totalheight=1.0in]{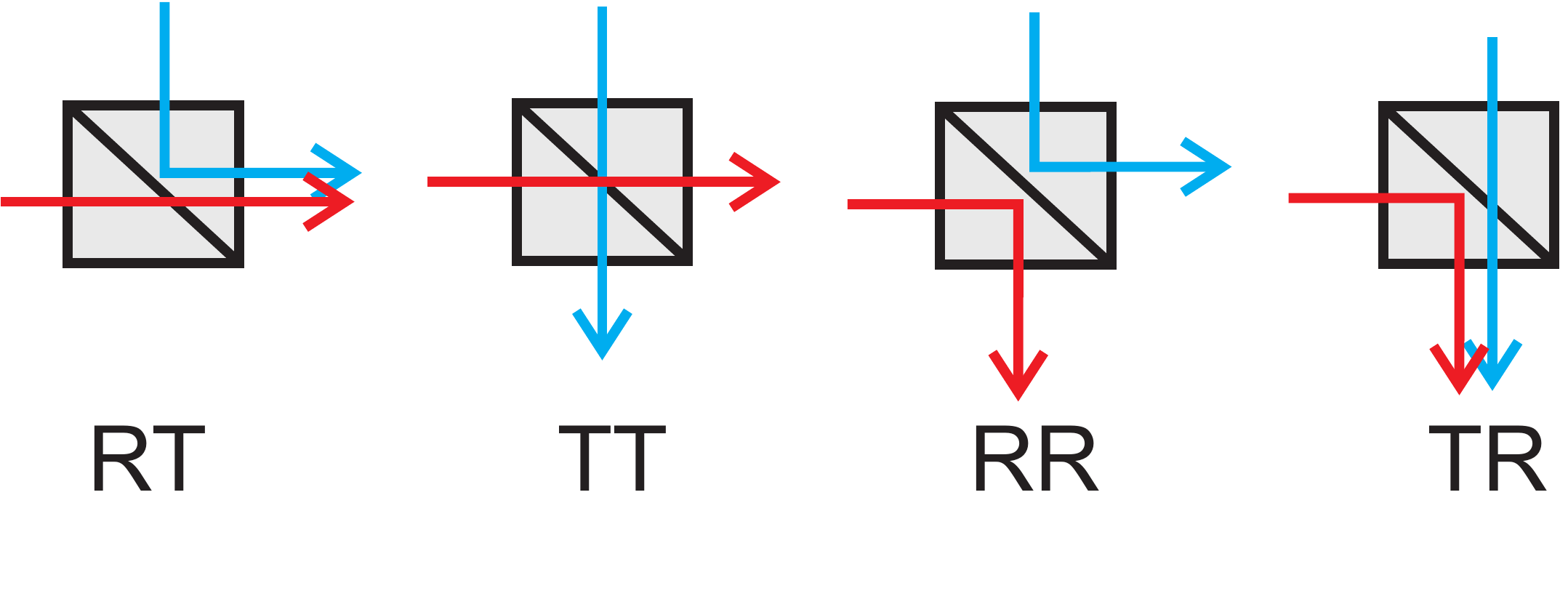}
\caption{The HOM effect: when two correlated photons are sent into different ports of a beam splitter, the middle two amplitudes destructively interfere, so that there are no coincidences detected at the two output ports. Even though the photons don't interact with each other, both of them always exit at the same port, but \emph{which} port is completely random. }\label{homfigfig}
\end{figure}

The archetypal example of a two-photon interference effect is the Hong-Ou-Mandel effect \cite{HOM}, shown schematically in Fig. \ref{homfigfig}. The two identical photons in this case are indistinguishable and are initially in a product state (non-entangled). They are sent into different ports of a beam splitter, and a coincidence measurement is made at the two output ports. Each photon can either be transmitted or reflected at the beam splitter, leading to four possible outcomes, as shown in the Figure. However, the two outcomes in the middle (both reflecting or both transmitting) are indistinguishable and appear with amplitudes of opposite signs, so they exhibit destructive interference. When the photons enter at widely separated times, they are distinguishable simply by their exit times, and so there is no interference. But as the delay $\tau$ decreases, they become less distinguishable and begin to destructively interfere. The indistinguishability is complete when $\tau=0$, and so at this point there are no coincidence counts: the two photons can still leave at either exit port, but they always exit at the \emph{same} randomly-chosen port, never at different ports. The result is the famous HOM dip \cite{HOM}, in which the coincidence rate dips sharply to zero as the time delay between the photons vanishes, and which has been verified experimentally in both photonic and atomic systems \cite{beugnon,wal,lopes}. For recent reviews of two-photon interference, see \cite{jach,bouchard,hoch}.

Note that here, even though the input is in a product state, what matters is that the \emph{output} state of the beam splitter is entangled: it is a superposition of two two-photon amplitudes $|\psi_a\rangle_3|\psi_b\rangle_3 + |\psi_a\rangle_4|\psi_b\rangle_4$, where $3$ and $4$ label the beam splitter output ports. The two coincidence or cross terms that could potentially arise in this superposition, $|\psi_a\rangle_3|\psi_b\rangle_4$  and $|\psi_a\rangle_4|\psi_b\rangle_3$, are indistinguishable by the detectors, and so are capable of canceling each other. The key point is indistinguishability of the interfering amplitudes \emph{at the detection stage}, regardless of the presence or absence of indistinguishability or entanglement at the input. In the setups described in subsequent sections, the distinguishability of multiple outcomes will effectively be erased by making it impossible for the detector to determine the path taken through the system by the photons, which then allows two-photon interference to take place.

Beam splitters are often an integral part of interferometers. But beam splitters are essentially one-way devices: the photons can never reverse direction to exit at the same port from which they entered. Recently, optical multiports have been studied \cite{simon1,osawa2,kim,osawa1} that do not share this directional bias: light entering a given port can exit at \emph{any} port, including the one through which it entered. The three-port versions of these unbiased multiports have been implemented in bulk devices in two different manners in \cite{osawa2} and \cite{kim}, while plans are currently underway to implement four-ports on integrated chips. The directionally-unbiased nature of these devices has implications for reducing the resources required in optical networks, as well as introducing new capabilities such as novel multiphoton interference effects, as has been demonstrated, for example, in \cite{simon2, osawa3, osawa4}. Further examples of the new capabilities introduced by these devices are discussed in this paper: in the following it will be shown that measurements with interferometers containing directionally-unbiased multiports can extract more information than the analogous standard beam splitter-based interferometers.

Of special interest is the directionally-unbiased four-port, which acts as a physical implementation of the four-dimensional Grover coin operation \cite{moore} that appears in studies of quantum walks \cite{carneiro}. When two correlated photons are input to such a four-port device, it has been shown \cite{simon2,osawa3} that higher-dimensional analogs of the HOM effect occur, in which both photons always exit at the same \emph{pair} of ports: either both reverse direction and exit at the two input ports, or they both continue in the forward direction and exit at the other two ports; there is never one photon reflecting back and one exiting in the forward direction. By introducing additional phase shifters this effect can be used to control the flow of two-photon entanglement through multiple branches of an optical network \cite{osawa4}.

In this paper, we demonstrate additional two-photon interference effects in interferometers that contain Grover four-ports. In the next section, the effect of these four-ports on two-photon amplitudes is reviewed, then in Section \ref{HOMsection} it is shown that a simple Grover multiport arrangement allows continuous interpolation between HOM dips (clustered, boson-like behavior) and anti-HOM peaks (anti-clustered, fermion-like behavior). Unlike previous approaches to demonstrating an anti-HOM effect, neither prior preparation of polarization-entangled states \cite{mechler} nor non-Hermiticity  \cite{li,vetlugin} in the system is required.  Switching between the different types of behavior can be done in real time simply by varying a phase shift. In Section \ref{gmz} it is shown that by joining two Grover multiports together it is possible to measure up to three separate phases simultaneously with a single interferometer that has the same topology as the Mach-Zehnder interferometer. This effect is then applied in Section \ref{groversagnac}, showing that a slight variation of the same setup can simultaneously measure rotations of the device about three orthogonal axes, providing a three-dimensional generalization of the standard Sagnac interferometer with a single device. Conclusions are briefly summarized in Section \ref{conclusionsection}.

Although we are not assuming in this paper that the input photons are entangled, we do assume that the input consists of temporally-coincident pairs of indistinguishable photons. Currently, the easiest way to produce such time-correlated photon pairs is using SPDC, so that the entanglement is already present for free, so to speak, without any additional effort. As is well-known, the use of entangled input states can lead to greatly enhanced resolution in imaging, interferometry, and lithography \cite{boto,israel,yurke,huver,untern}, though at the expense of increased fragility against environmental disturbances. Particularly relevant to the proposed setup in this paper, entanglement can improve the resolution of Sagnac interferometry \cite{fink}. So the setups described below are naturally well-suited to entangled-photon generalizations. But we leave the study of entanglement-based enhancements for study elsewhere, and simply assume the use here of classically time-correlated pairs of indistinguishable photons as input. Throughout this paper, ``correlated photons'' will always mean two indistinguishable photons arriving in a time-correlated pair, and will not necessarily imply entanglement.

\section{Grover four-ports}\label{groversection}

\begin{figure}
\centering
\includegraphics[totalheight=1.6in]{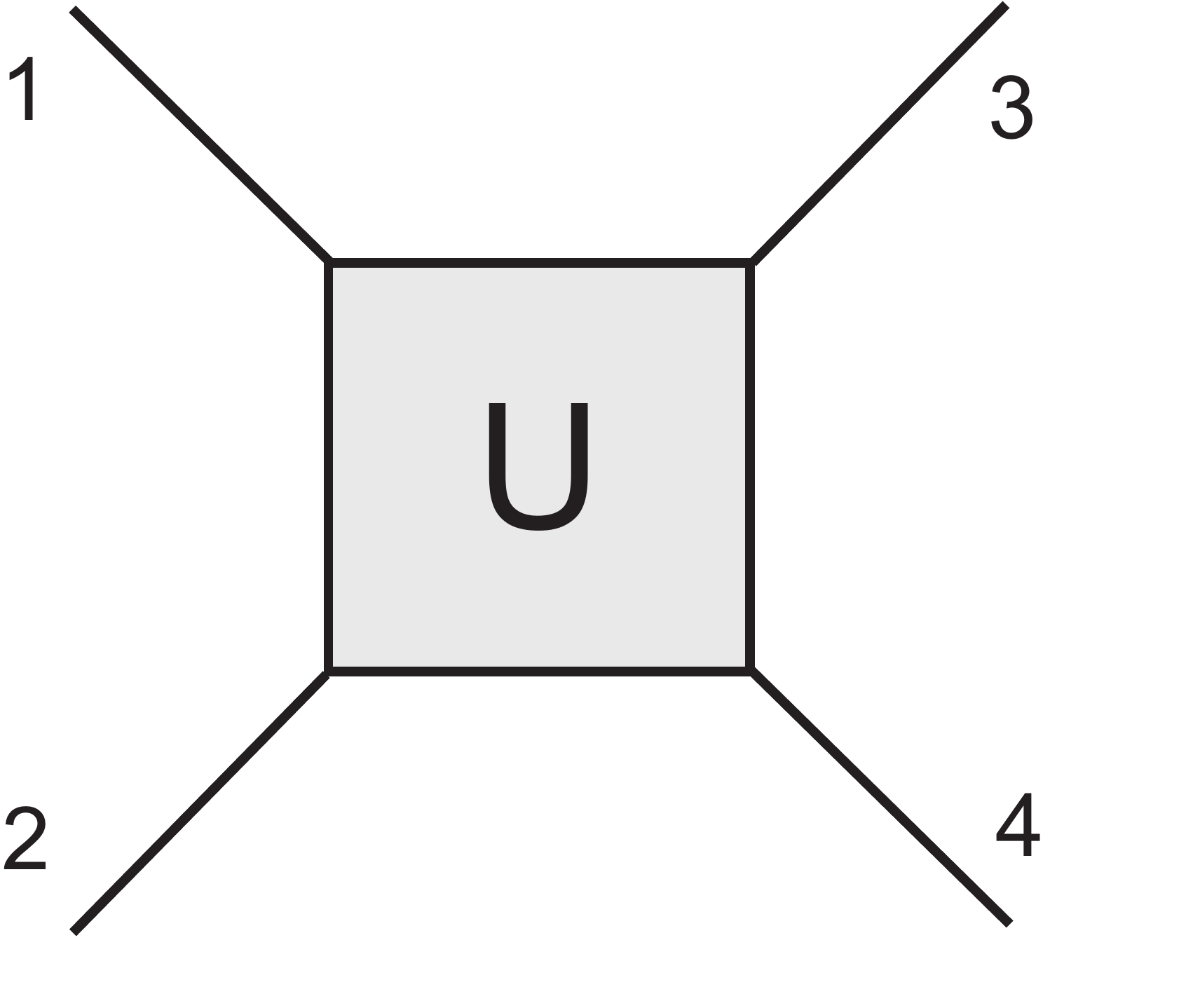}
\caption{The Grover optical multiport converts input at the four ports into output amplitudes at all four ports via unitary four-by-four matrix $U$. }\label{groverfig}
\end{figure}

The Grover coin operator, originally introduced for quantum walk applications, has a four-dimensional version that takes the matrix form:
\begin{equation}U={1\over 2}\left( \begin{array}{cccc} -1 & 1 & 1 & 1 \\ 1 & -1 & 1 & 1 \\ 1 & 1 & -1 & 1 \\ 1 & 1 & 1 & -1\end{array}\right) .\end{equation} A recent recipe for physically implementing directionally-unbiased $n$-ports \cite{simon1} includes the four-dimensional Grover coin matrix as a special case for $n=4$. In this case, the rows and columns represent the amplitudes entering and exiting from the four optical input/output ports of the device, so that $U$ represents the evolution matrix from the input optical state before encountering the multiport to the output state after the multiport, $|\psi\rangle_{out} = U|\psi\rangle_{in} $. Labeling the four ports as shown in Fig. \ref{groverfig}, the optical states are of the form \begin{equation}|\psi\rangle = \left( \begin{array}{c} \phi_1 \\ \phi_2 \\ \phi_3 \\ \phi_4  \end{array}\right), \end{equation} where $\phi_i$ is the amplitude at port $i$.  Here, we are not explicitly including directional labels on the states, indicating whether the photon amplitude is moving inward or outward motion the multiport. In what follows, we will continue to supress these directional labels, but it should be understood that the direction flips from inward to outward at each multiport encounter.

Suppose that a pair of correlated photons are input, one on each of two lines of the Grover multiport (at lines 1 and 2 of Fig. \ref{groverfig}, say). Then equal amplitudes appear for a transmitting state $|\psi_t\rangle$ which exits out the opposite pair of ports (3 and 4) and a state $|\psi_r\rangle$ which reflects back out the original two ports (1 and 2). Here, $|\psi_t\rangle$  and $|\psi_r\rangle$, defined below, are states that propagate unchanged during encounters with the multiports, aside from a reflection (in the case of $|\psi_r\rangle$) or a translation from one side of the multiport to the other (in the case of $|\psi_t\rangle$). Switching to notation that will be more convenient here, let $|n_1 n_2 n_3 n_4\rangle$ denote the state with $n_j$ photons on the $j$th line. Let $\hat a_j^\dagger$ be the creation operator for a photon on the $j$th line, so that \begin{eqnarray} & & |n_1,n_2,n_2,n_3\rangle \\
& & \qquad  = {1\over \sqrt{n_1 !n_2!n_3!n_4!}}(a_1^\dagger)^{n_1} (a_2^\dagger)^{n_2} (a_3^\dagger)^{n_3} (a_4^\dagger)^{n_4}  |0\rangle.\nonumber\end{eqnarray} Then
the input state with one photon on each of two lines, \begin{equation}|\psi\rangle_{in}=|1100\rangle\label{psiin}\end{equation} splits into a superposition of two states, \begin{equation}|\psi\rangle_{in}\to |\psi\rangle_{out} = {1\over \sqrt{2}}\left( |\psi_t\rangle +|\psi_r\rangle \right),\end{equation} where the arrow indicates one time step of evolution, taking the state before entering the multiport into the state exiting it. Aside from the reversal of direction (ingoing to outgoing), and of a translation one step to the right in the case of $|\psi_t\rangle$, these states are unchanged by the multiport: \begin{eqnarray}|\psi_t\rangle &=&{1\over 2} \left( |0020\rangle + |0002\rangle \right)  + {1\over \sqrt{2}}|0 011\rangle \\ & =& {1\over {2\sqrt{2}}} \left( \hat a_3^\dagger +\hat a_4^\dagger \right)^2 |0\rangle \end{eqnarray} continues to be transmitted unchanged each time it encounters a multiport, while the state \begin{eqnarray}|\psi_r\rangle &=& -{1\over 2} \left( |2000\rangle + |0200\rangle \right)+{1\over \sqrt{2}}|1100\rangle \\ &=& -{1\over {2\sqrt{2}}} \left( \hat a_1^\dagger -\hat a_2^\dagger \right)^2 |0\rangle \end{eqnarray} simply reflects backward at each multiport encounter \cite{simon2}.

\begin{figure}
\begin{center}
\subfigure[]{
\includegraphics[height=2.0in]{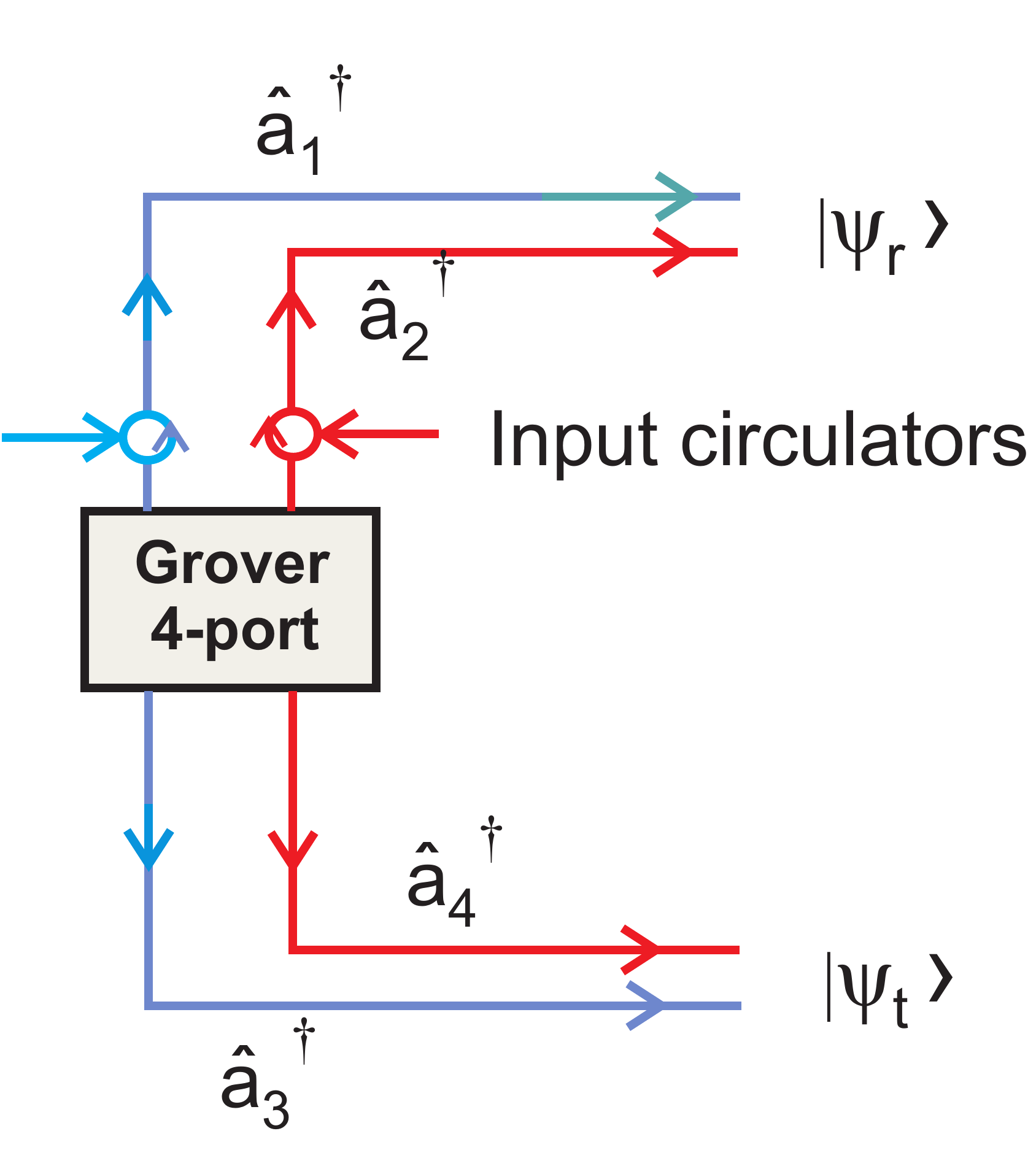}\label{multioutfig}}\quad
\subfigure[]{
\includegraphics[height=2.0in]{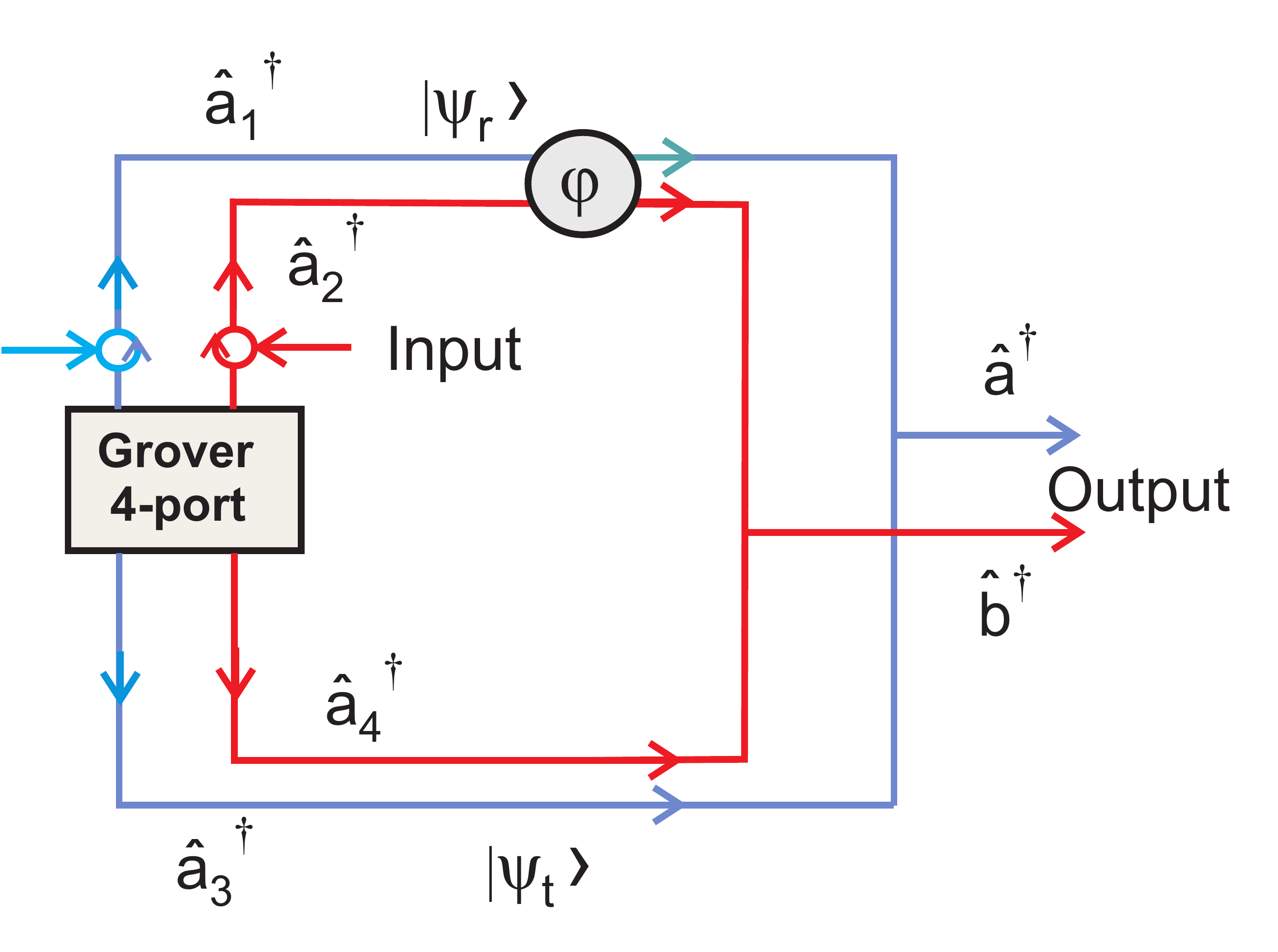}\label{phaseshiftfig}}\quad
\caption{(a) Optical circulators are used to input two photons into the multiport, via lines 1 and 2. The multiport produces an output that is a 50-50 superposition of two two-photon amplitudes: $|\psi_t\rangle $ exits on lines 3 and 4, while $|\psi_r\rangle $ exits on lines 1 and 2.  (b) To produce new interference effects, the lines 2 and 4 are merged, as are the lines 1 and 3. A common phase shift has been added to lines 1 and 2, relative to 3 and 4.}\label{multisetupfig}
\end{center}
\end{figure}

So, sending two photons into the multiport via optical circulators, the output has equal amplitudes for states $|\psi_t\rangle$ and $|\psi_r\rangle$. Until a measurement is made, each of the two two-photon amplitudes exists simultaneously, moving in opposite directions, as shown in Fig. \ref{multioutfig}. In other words, the two-photon state exiting the multiport is now entangled in the spatial degree of freedom.

\section{Tuning between HOM and anti-HOM interference effects}\label{HOMsection}

We now continue to consider a multiport with two input photons inserted via optical circulators. As in the previous section, this results in a superposition of a state $|\psi_r\rangle$ exiting at the top two ports and $|\psi_t\rangle$ at the bottom two.
But, suppose now that pairs of amplitudes are brought together and allowed to interfere. Start by imposing a phase shift $\phi$ on each of the photons in the top line ($|\psi_r\rangle \to e^{2i\phi}|\psi_r\rangle$), and then merge the blue $\hat a_1$ and $\hat a_3$ lines. Merging here means that they both are directed to a common detector that can't distinguish whether the detected photon came from port $1$ or port $3$. Both modes trigger the same detection event at the end. As a practical matter, we can then effectively replace both $\hat a_1^\dagger$ and $\hat a_3^\dagger$ by a single operator representing the mode that the detector sees (call it $\hat a^\dagger$) . This is represented schematically by the merger of the blue lines in Fig. \ref{phaseshiftfig}. Similarly, combine the two red $\hat a_2$ and $\hat a_4$ lines at a single detector, effectively replacing both $\hat a_2^\dagger$ and $\hat a_4^\dagger$ by a single detected mode $\hat b^\dagger$. The desired merger of the beams can be accomplished by means of a beam splitter, although this would entail the loss of some light out the unused ports. A better approach is to merge them via evanescent coupling within a fiber Y-coupler, which can be designed to transfer nearly all of the light from the two incoming paths into a single output path with minimal loss. Or, if working in free space, mirrors can simply be used to direct the two beams to the same detector, since all that is needed is indistinguishability between the two paths the photons could take to each detector.

%

Then at the $a$ and $b$ ports on the right, the output state will be:
\begin{equation}|\psi_{out}\rangle =  e^{i\phi} \left[ -{i\over \sqrt{2}} \left( |02\rangle +|20\rangle \right)\sin \phi +|11\rangle \cos\phi \right] ,\end{equation} where $|n,m\rangle$ now means the state with $n$ photons in the $a$ line and $m$ photons in the $b$ line.

For the special case of $\phi={\pi\over 2}$, we find \begin{equation}|\psi_{out}\rangle = {1\over \sqrt{2}}\left( |02\rangle +|20\rangle \right).\end{equation} In other words, we find the usual HOM effect, with an entangled state of two photons either in one line or in the other, and \emph{no} coincidences between the $a$ and $b$ lines.

On the other hand, if we set $\phi =0$, then we get an anti-HOM effect, in which we have a product state containing \emph{only} coincidences: \begin{equation}|\psi_{out}\rangle = |11\rangle .\end{equation} There is now a coincidence \emph{peak} at zero delay, instead of a dip.

So, by turning a dial to change the value of $\phi$ we can continuously, and in real time, interpolate between \emph{no} coincidences and \emph{all} coincidences, or between maximal spatial-mode entanglement and a product state.  Unlike previous methods for interpolating between HOM and anti-HOM effects, there is no need for loss or non-unitary evolution to be imposed \cite{li,vetlugin}, and there is no need for different polarization-entangled states to be prepared for each type of behavior \cite{mechler}. The same interferometer and light source can be toggled easily between the two types of behavior without reconfiguring the setup.

The source of the effect is easy to see. After merging the pairs of lines, the outputs of the top pair and of the bottom pair are, respectively,
\begin{equation}-e^{2i\phi} (\hat a^\dagger -\hat b^\dagger )^2 |0\rangle  \quad \mbox{and} \quad \hat (a^\dagger +\hat b^\dagger)^2 |0\rangle .\end{equation}
Adding the two terms with $\phi={\pi\over 2}$, the cross terms cancel and we get zero coincidence:
\begin{equation} \left[ (\hat a^\dagger -\hat b^\dagger )^2  + (\hat a^\dagger +\hat b^\dagger)^2\right] |0\rangle =2 (\hat a^{\dagger 2}+\hat b^{\dagger 2}) |0\rangle.\end{equation} But for $\phi=0 $, the squared terms cancel, leaving \emph{only} coincidences:  \begin{equation} \left[ - (\hat a^\dagger -\hat b^\dagger )^2  + (\hat a^\dagger +\hat b^\dagger)^2 \right] |0\rangle = 4\hat a^\dagger \hat b^\dagger  |0\rangle .\end{equation}


%

The continuous tuning between HOM and anti-HOM can be made clear experimentally by introducing several time delays into the system.
Imagine, for example, $\tau_1$ and $\tau_2$ are delays between the two lines within a single two-photon amplitude; $\tau_0$ is a delay between one two-photon amplitude and the other. The phase shifts are now due to the time delays, $\phi_j=\omega \tau_j$, where $\omega$ it the common frequency of the indistinguishable photons.

\begin{figure}
\centering
\includegraphics[totalheight=2.0in]{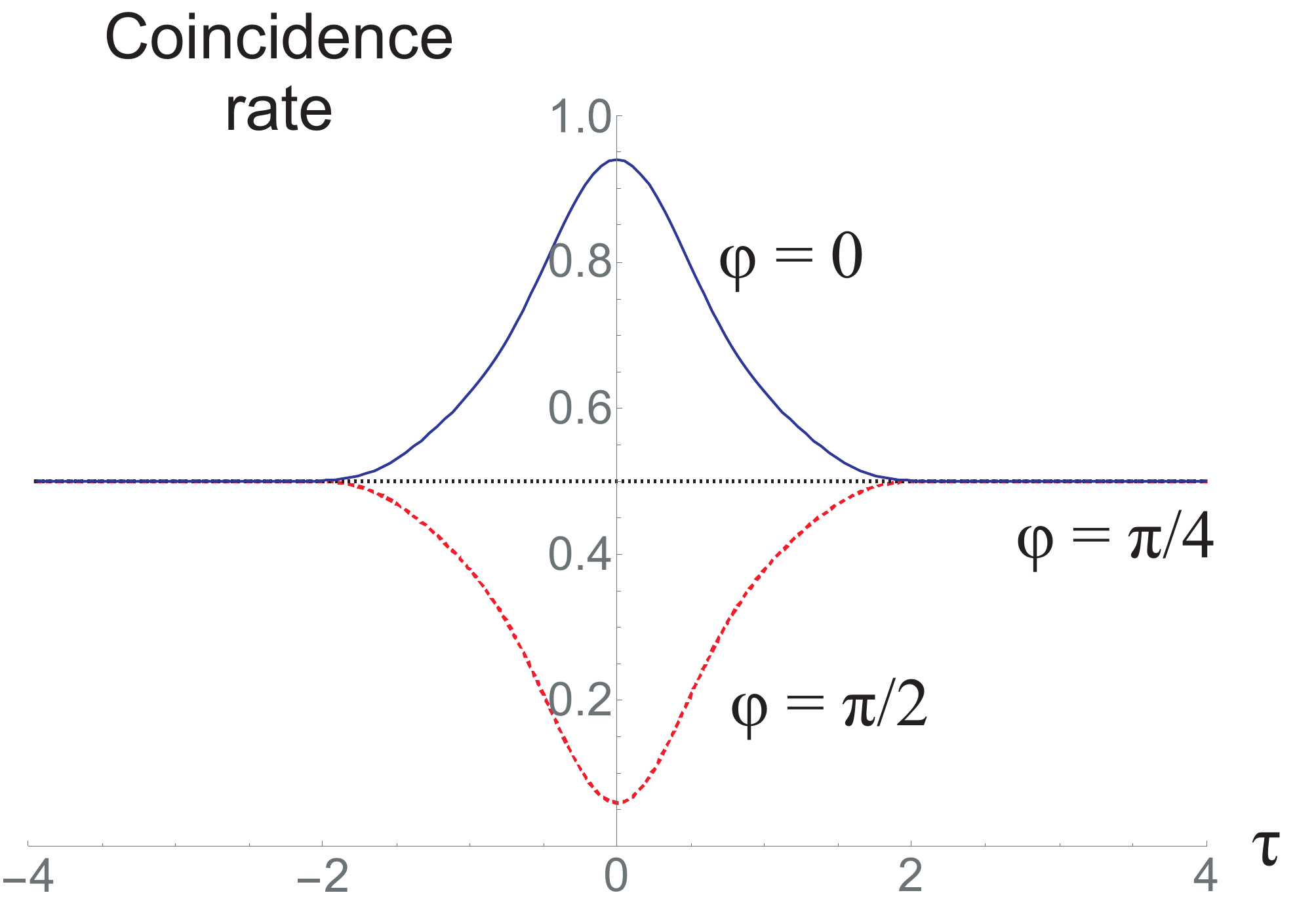}
\caption{By tuning
the value of $\phi$, the coincidence count can be made to exhibit boson-like behavior (an HOM dip for $\phi={\pi\over 2}$, red, dashed curve), classical-like behavior (no two-photon interference, for $\phi ={\pi\over 4}$, black, dotted line), or anti-clustered fermion-like behavior (a coincidence peak, for $\phi = 0$, blue, solid curve). The plots here are for scans over $\tau_0=\phi/\omega$, but scans over $\Delta \tau =\tau_2-\tau_1$ show similar behavior. The simulations shown here use a spectral amplitude $\Phi(\omega )$ proportional to a $sinc$ function. }\label{phidelayfig}
\end{figure}

Assuming that all of the photons have the same spectral amplitude, $\Phi(\omega )$, a little work leads to the coincidence probability between the two outputs:
\begin{eqnarray}p&=&{1\over 2} +{1\over 2}\cos 2\phi \int d\omega_a\; d\omega_b |\Phi (\omega_a)\Phi (\omega_b)|^2 \nonumber \\
& & \qquad\qquad\qquad\qquad\times \cos \left[ \omega_b \Delta \tau +(\omega_a +\omega_b)\tau_0 \right]\nonumber \\ & &\qquad
+{1\over 2}\sin 2\phi \int d\omega_a\; d\omega_b |\Phi (\omega_a)\Phi (\omega_b)|^2 \nonumber \\
& & \qquad\qquad\qquad\times\sin \left[ -\omega_b \Delta \tau +(\omega_a -\omega_b)\tau_0 \right]
,\end{eqnarray} where $\Delta \tau = \tau_2-\tau_1$. Imagine holding either $\tau_0$ constant while varying $\Delta \tau$, or else holding $\Delta \tau $ fixed while varying $\tau_0$. In either case, the result is an HOM dip or anti-HOM peak with a width similar to the standard HOM dip for the same assumed $\Phi(\omega)$ (Fig. \ref{phidelayfig}). But notice that in the first case (varying $\Delta \tau$), the dip doesn't measure the delays between photons within the same input pair; rather, it measures the \emph{difference} in delays between two different pairs of paths. Effectively it measures the \emph{gradient} of the inter-photon delay between the pairs in the upper and lower two-photon paths. In other words, if $\Delta \phi_r$ is the difference between photons in the upper two lines (in the reflected amplitude) and $\Delta \phi_t$ is the difference between the two photons in the lower path (in the transmitted amplitude), then the dip is measuring the difference of these two differences, $\Delta^2 \phi\equiv \Delta \phi_r-\Delta\phi_t$.

\section{Multiple simultaneous phase measurements: the Grover-Mach-Zehnder Interferometer}\label{gmz}
\vskip 5pt

\begin{figure}
\begin{center}
\subfigure[]{
\includegraphics[height=1.8in]{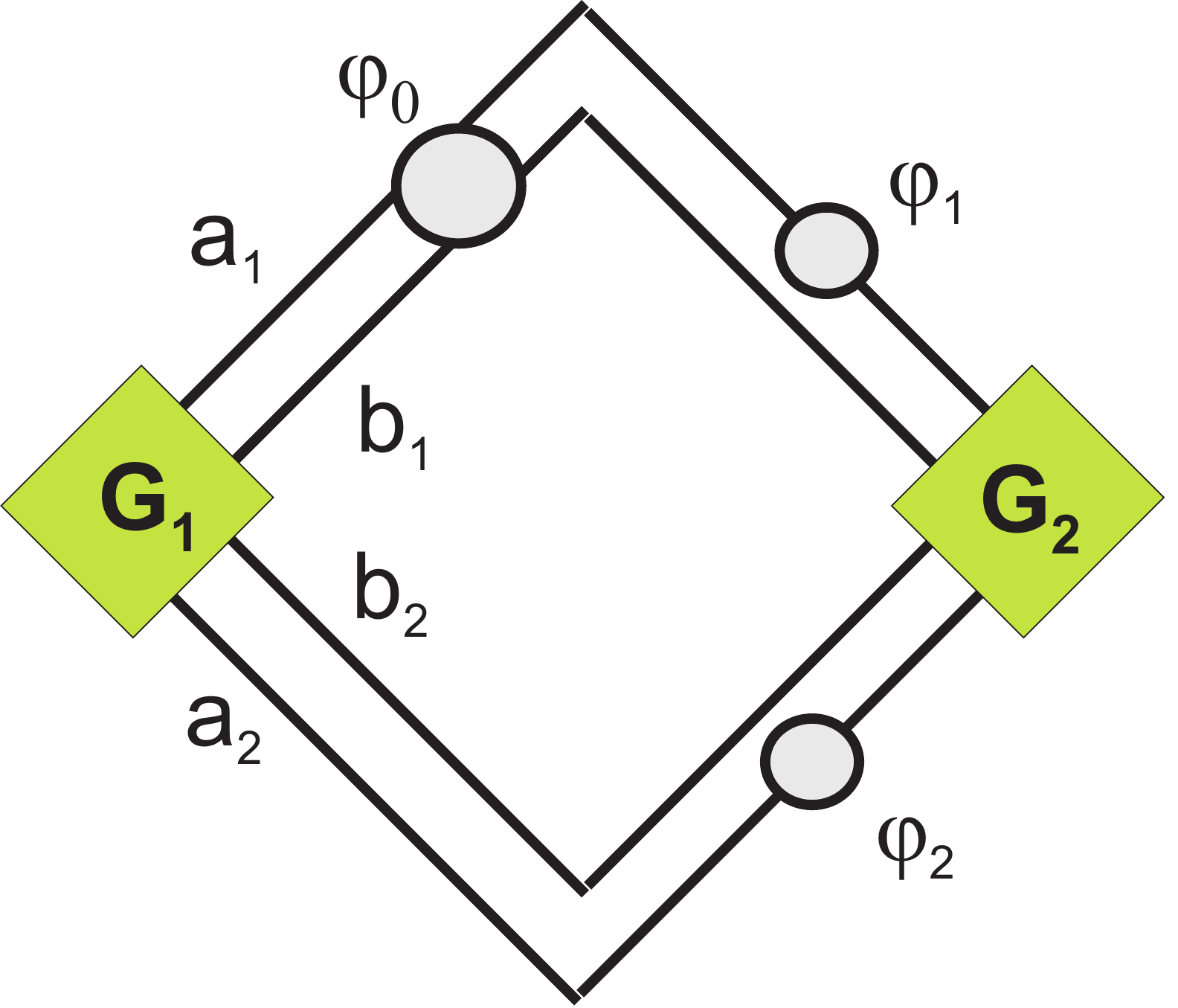}\label{MZafig}}\qquad \qquad\qquad
\subfigure[]{
\includegraphics[height=1.8in]{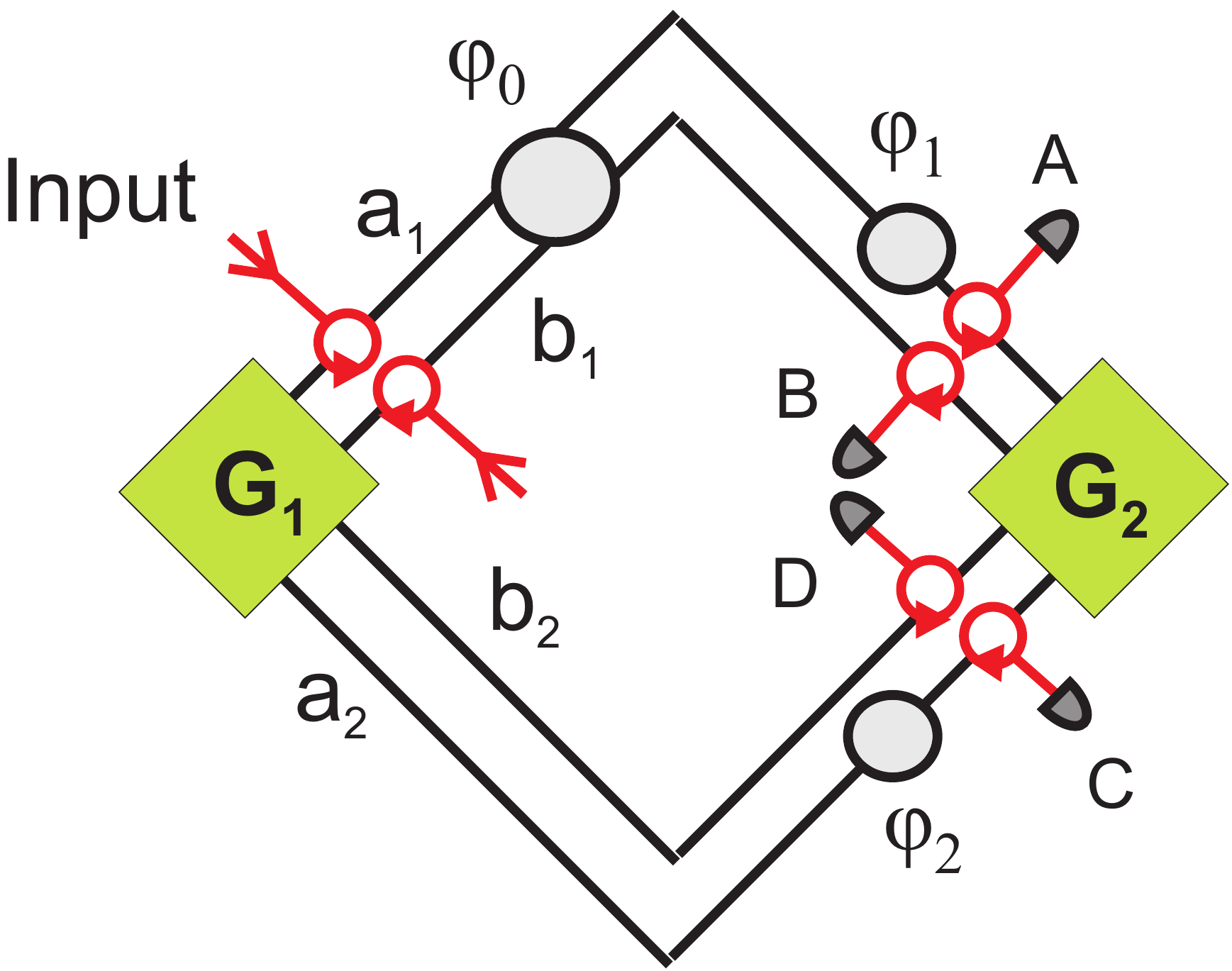}\label{MZbfig}}
\caption{(a) Mach-Zehnder interferometers can be generalized, replacing the beam splitters by Grover four-ports.  The two four-ports are connected by two lines in the upper branch and two in the lower branch. Phase shifts (represented by the grey-shaded circles) are added as shown. (b) Photons are coupled in and out of the system with circulators (the red circles). The output lines are fed to four detectors, labeled A, B, C, D. } \label{walkfig}
\end{center}
\end{figure}

The effect above results from placing a Grover four-port into an interferometer. This naturally leads to the question of what other effects might be possible by building other interferometers from four-ports. In Fig. \ref{MZafig}, a Grover four-port analog of the Mach-Zehnder interferometer is shown. A two-photon state of the form of Eq. \ref{psiin} is input to the left-hand four-port ($G_1$), which splits it into two two-photon amplitudes travelling along the upper and lower branches to the second four-port, $G_2$. The photons are coupled into $G_1$ by a pair of circulators (the red circles in Fig. \ref{MZbfig}) at the left, and at the right output is extracted by additional circulators coupled to four detectors (A, B, C, D). Tagging photon arrival at each detector with a time stamp, we can measure the coincidence rates $R_{AB}, R_{AC}$, etc. between any pair of detectors. There will also be events where two photons arrive at the same detector, but in what follows we are only interested in the coincidence events.

Suppose three phase shifts are added as in Fig. \ref{walkfig}. $\phi_0$ is the shift between upper and lower branches (specifically, the shift of $b_2$  relative to $b_1$), while $\phi_1 $ and $\phi_2$ are the relative phase shifts between the two lines within each of those branches (between $b_1$ and $a_1$ in the first case, and between $b_2$ and $a_2$ in the second). So, for example, the phase of $a_1$ is shifted higher than that of $b_2$ by $\phi_0+\phi_1$.

Some of the coincidence rates are equal ($R_{AC}=R_{BD}$ and $R_{AD}=R_{BC}$), so $R_{BD}$ and $R_{BC}$ no longer need to be considered. That leaves four independent coincidence rates, which are readily calculated:

\begin{eqnarray}
R_{AC}&=& R_{BD}\nonumber \\
&=& R_0 \left| e^{i(2\phi_0+\phi_1)}\sin\phi_1-e^{i\phi_2}\sin\phi_2\right|^2 \nonumber \\ &=& R_0 \left[ \sin^2\phi_1+\sin^2\phi_2\nonumber \right.\\ & & \quad -\left.2\sin\phi_1\sin\phi_2\cos(2\phi_0+\phi_1-\phi_2)\right]\\
R_{AD}&=& R_{BC}\nonumber \\
&=& R_0 \left|e^{i(2\phi_0+\phi_1)}\sin\phi_1+e^{i\phi_2}\sin\phi_2\right|^2 \nonumber \\ &=& R_0 \left[ \sin^2\phi_1+\sin^2\phi_2 \right. \nonumber \\ & &\left.+2\sin\phi_1\sin\phi_2\cos(2\phi_0+\phi_1-\phi_2)\right]\\
R_{AB}&=& R_0 \left|e^{i(2\phi_0+\phi_1)}(\cos\phi_1+1)+e^{i\phi_2}(\cos\phi_2+1)\right|^2 \nonumber \\ &=& R_0 \left[ (\cos\phi_1+1)^2+(\cos\phi_2+1)^2\right. \\ & & \left. +2(\cos\phi_1+1)(\cos\phi_2+1)\cos(2\phi_0+\phi_1-\phi_2)\right] \nonumber \\
R_{CD}&=& R_0 \left|e^{i(2\phi_0+\phi_1)}(\cos\phi_1-1)+e^{i\phi_2}(\cos\phi_2-1)\right|^2 \nonumber \\ &=& R_0 \left[ (\cos\phi_1-1)^2+(\cos\phi_2-1)^2\right. \\ & & \left. +2(\cos\phi_1-1)(\cos\phi_2-1)\cos(2\phi_0+\phi_1-\phi_2)\right] \nonumber ,
\end{eqnarray} where $R_0$ is a fixed constant determined by the rate at which photons are entering the initial multiport. Once these coincidence rates are measured, the formulas can be inverted to solve for the four unknowns, $R_0$, $\phi_0$, $\phi_1$, $\phi_2$.

Simplifications can be obtained by taking sums and differences of rates:
\begin{eqnarray}{{R_{AC}+R_{AD}}\over {R_0}} &=& 2 (\sin^2\phi_1 +\sin^2\phi_2) \label{coa}\\
{{R_{AD}-R_{AC}}\over {R_0}} &=& 4 \sin\phi_1 \sin\phi_2 \cos(2\phi_0 +\phi_1-\phi_2) \label{cob}\\
{{R_{AB}+R_{CD}}\over {R_0}} &=& 2(1+\cos^2\phi_1) +2(1+\cos^2\phi_2)\nonumber \\  & & \qquad +4\cos(2\phi_0 +\phi_1-\phi_2)\nonumber \\ & & \qquad\qquad \times (1+\cos\phi_1\cos\phi_2)\label{coc}\\
{{R_{AB}-R_{CD}}\over {R_0}} &=& 4(\cos\phi_1+\cos\phi_2)\nonumber \\
& & \qquad\times [1+\cos(2\phi_0 +\phi_1-\phi_2)] ,\label{cod}
\end{eqnarray} where $R_0$ is some constant that will depend on the rate at which photons enter the system.


This set of four independent equations allows the three phase shifts $\phi_0, \phi_1, \phi_2$ and the constant $R_0$ to be determined from the experimentally-measured coincidence rates.  They can be readily solved numerically for a given set of measured coincidence rates.

In practice, $R_0$ can be measured independently for the given source (it is essentially the total down conversion rate). So, if it is assumed that $R_0$ is known, then in fact the equations simplify to the point where it is possible to give exact solutions; Mathematica, for example, can solve to provide a set of exact analytic solutions. However, these solutions are more than a page long when written out, and so we do not choose to reproduce them here. But, despite their complexity, the exact solutions could be used instead of numerical solutions, meaning that the only errors would be due to imprecision in the measured data.

\begin{figure}
\centering
\includegraphics[totalheight=1.6in]{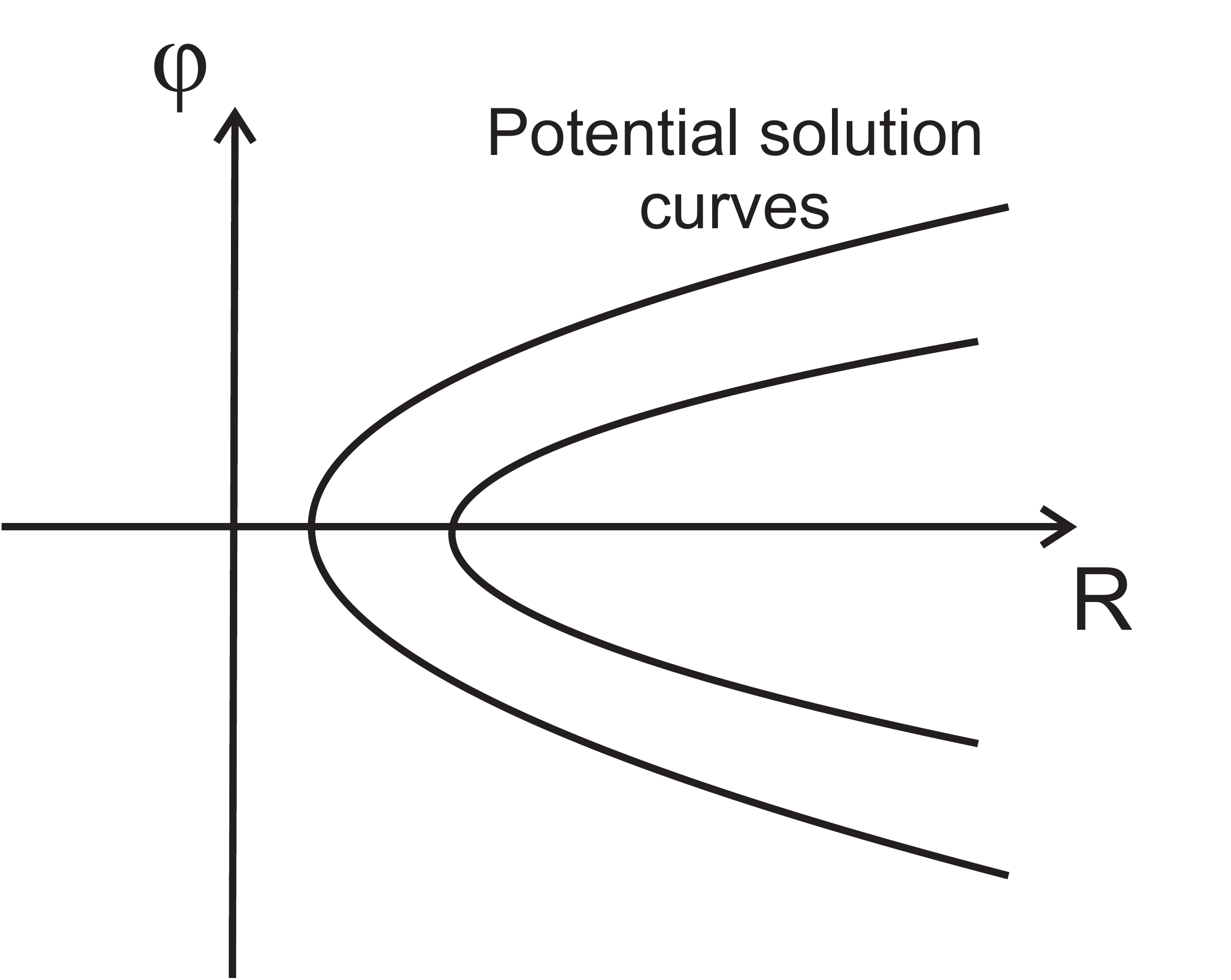}
\caption{Solving for the phases as functions of the measured coincidence rates gives multiple possible solution surfaces, depicted schematically in this figure as curves. Calibration of the system before use by measuring the coincidence rates at zero phase shifts allows determination of which solution surface the system lies upon. However, in most cases, this will still leave a sign ambiguity (between the upper and lower branches of the surface), allowing only the absolute value of the phase to be determined in these cases. }\label{solutionfig}
\end{figure}

A more serious issue is that, because of the squares in the equations, there are multiple solutions for the phases at a given set of coincidence rates, and so the correct solution for the given physical problem must be singled out from among the mathematical possibilities. In practice, this can be done by carrying out an initial calibration before the system is used: measure all the coincidence rates for the given source when the system has no phase shifts introduced. Only one of the multiple solution surfaces will pass through the correct coincidence values. See Fig. \ref{solutionfig} for a simplified picture of this. This uniquely determines the solution curve, which in turn determines the phase up to an overall sign.

In general, there could still be overall sign ambiguities left over in the phases (the choice between upper and lower branch of each curve in Fig. \ref{solutionfig}) after determining the correct solution curve. Note that these sign ambiguities often don't appear in simplified cases where only two phase shifts are independent; this can be seen in the example given below.

The end result is that with a Grover four-port generalization of the Mach-Zehnder interferometer, \emph{measurements of the coincidence rates allow for the simultaneous determination of the magnitudes (although possibly not the signs) of three phase shifts with a single interferometer}.
It should also be pointed out that because of the factors of 2 that arise in the sines and cosines of Eqs. \ref{coa}-\ref{cod}, the resolution with which $\phi_0$ is determined will be double the resolution of the usual Mach-Zehnder.

\vskip 5pt
%

In special cases where there are only two independent phases, a simple analytic solution is readily obtained. For example, suppose that $\phi_1=\phi_2$, leaving $\phi_0$ and $\phi_1$ to be determined.  It is straightforward to show that:
\begin{eqnarray}\cos 2\phi_0 &=& \Lambda_1 \\
\cos\phi_1 &=& {{1\pm \sqrt{1-\Lambda_1^2}}\over {\Lambda_2}} ,\end{eqnarray}
where \begin{eqnarray}
\Lambda_1&=& {{R_{AD}-R_{AC}}\over {R_{AD}+R_{AC}}}\\
\Lambda_2&=&  {R_{AB}-R_{CD}}\over {R_{AB}+R_{CD}} .
\end{eqnarray}
Since these equations are linear, the aforementioned sign ambiguities do not occur.

\begin{figure}
\centering
\includegraphics[totalheight=1.8in]{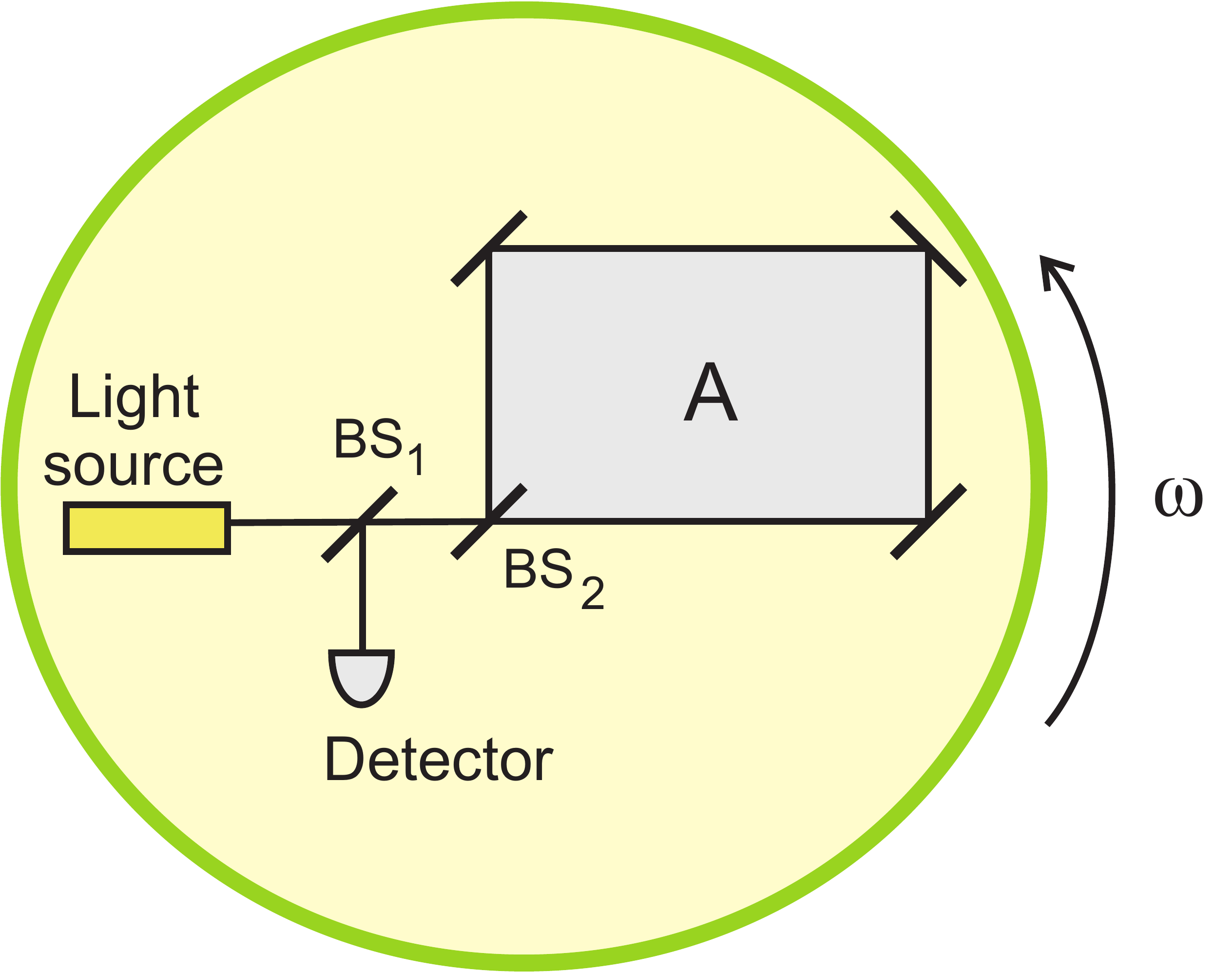}
\caption{Schematic of the standard Sagnac interferometer. As the apparatus rotates, the detector sees an interference pattern due to the fact that the two counter-propagating beams gain a relative phase shift $\delta \phi$ that is proportional to the rotational speed $\omega$. }\label{sagnacfig}
\end{figure}

\begin{figure}[t!]
\centering
\includegraphics[totalheight=1.8in]{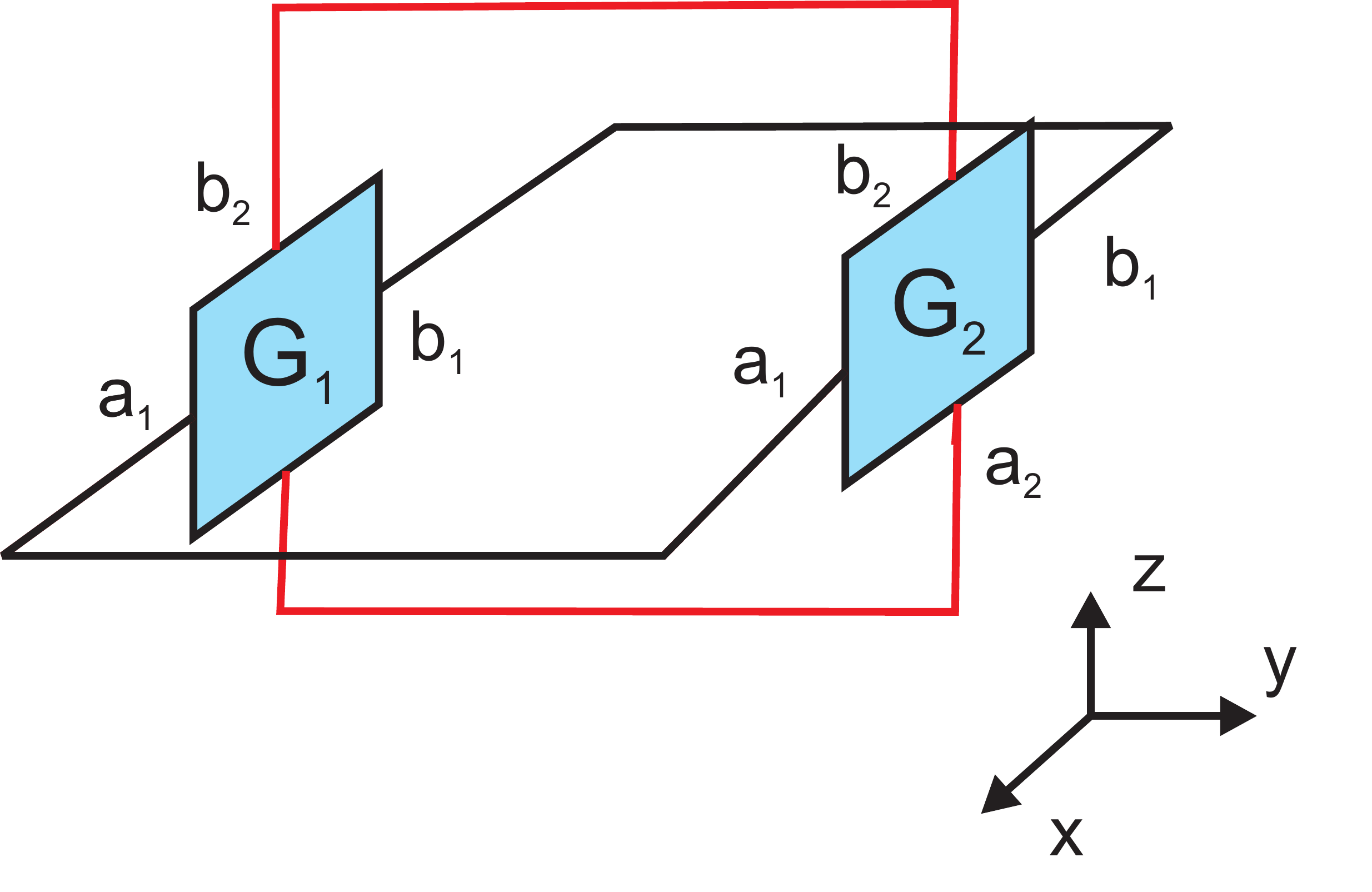}
\caption{The Grover-Mach-Zehnder interferometer of Fig. \ref{walkfig} can be put into a non-planar configuration, allowing rotations about two axes to be measured via the coincidence rates.  Although not drawn in, detectors are coupled in via circulators after $G_2$ as in Fig. \ref{walkfig}.}\label{2Dsagfig}
\end{figure}

\pagebreak
\section{The Grover-Sagnac interferometer}\label{groversagnac}
\vskip 5pt

The well-known Sagnac effect \cite{sag,post,log,hari} allows rotation rates to be measured via interference, and has a number of applications to fiber-optical gyroscopes \cite{herve}, measurement of gravitational effects \cite{sun}, and quantum sensing \cite{paiva}. Sagnac-based system have even been devised \cite{rest,matos} to test quantum effects such as HOM interference and the Aharonov-Bohm effect in non-inertial frames. A schematic depiction of the standard Sagnac interferometer is shown in Fig. \ref{sagnacfig}.
The incident beam is split into two parts by $BS_1$, and these parts propagate around the loop in opposite directions. If the loop is rotating at angular speed $\omega$
then the beams in the two directions take different times to make a complete loop and reach the detector. The time difference is readily shown to be
\begin{equation}\delta t ={{4A\omega}\over {c^2-r^2\omega^2}} \approx {{4A\omega}\over {c^2}},\end{equation} where $A$ is the area enclosed by the loop.
This leads to a phase difference of
\begin{equation}\delta \phi =\omega\; \delta t \approx {{8\pi A\omega }\over {\lambda c}}\label{phasearea}\end{equation} between the two counter-propagating beams. Since $\delta \phi$ is proportional to $\omega$, measurement of the phase shift amounts to a measurement of the rotation rate about the axis perpendicular to the plane of the interferometer.

A variation on the ideas of the previous section allows simultaneous measurement of rotations about three orthogonal axes with a single interferometer.
Consider measurements about just two axes first. The Grover-Mach-Zehnder interferometer above can be configured so that $a_1$ and $b_1$ lines form a loop in the $xy$-plane, while the $a_2$ and $b_2$ lines form a loop in the $yz$-plane. Schematically (leaving out the circulators needed to couple in the input), this is shown in Fig. \ref{2Dsagfig}.

Then the phase shifts $\phi_1$ and $\phi_2$ of the previous section are caused by rotations about the $z$- and $x$-axes, respectively, as in a pair of standard Sagnac interferometers:
\begin{eqnarray} \phi_1 &=& \phi_z \; =\; {{8\pi}\over {\lambda c}}\omega_z A_z \\
\phi_2 &=& \phi_x\; =\; {{8\pi}\over {\lambda c}}\omega_x A_x ,
\end{eqnarray}
where $\omega_j$ and $A_j$ are the frequencies of rotations about the $j$th axes and the areas enclosed by the corresponding loops. If a loop is not confined to a single plane, then the relevant area is the area of the loop's projection onto the plane perpendicular to the corresponding rotation axis.

\begin{figure}
\centering
\includegraphics[totalheight=2.2in]{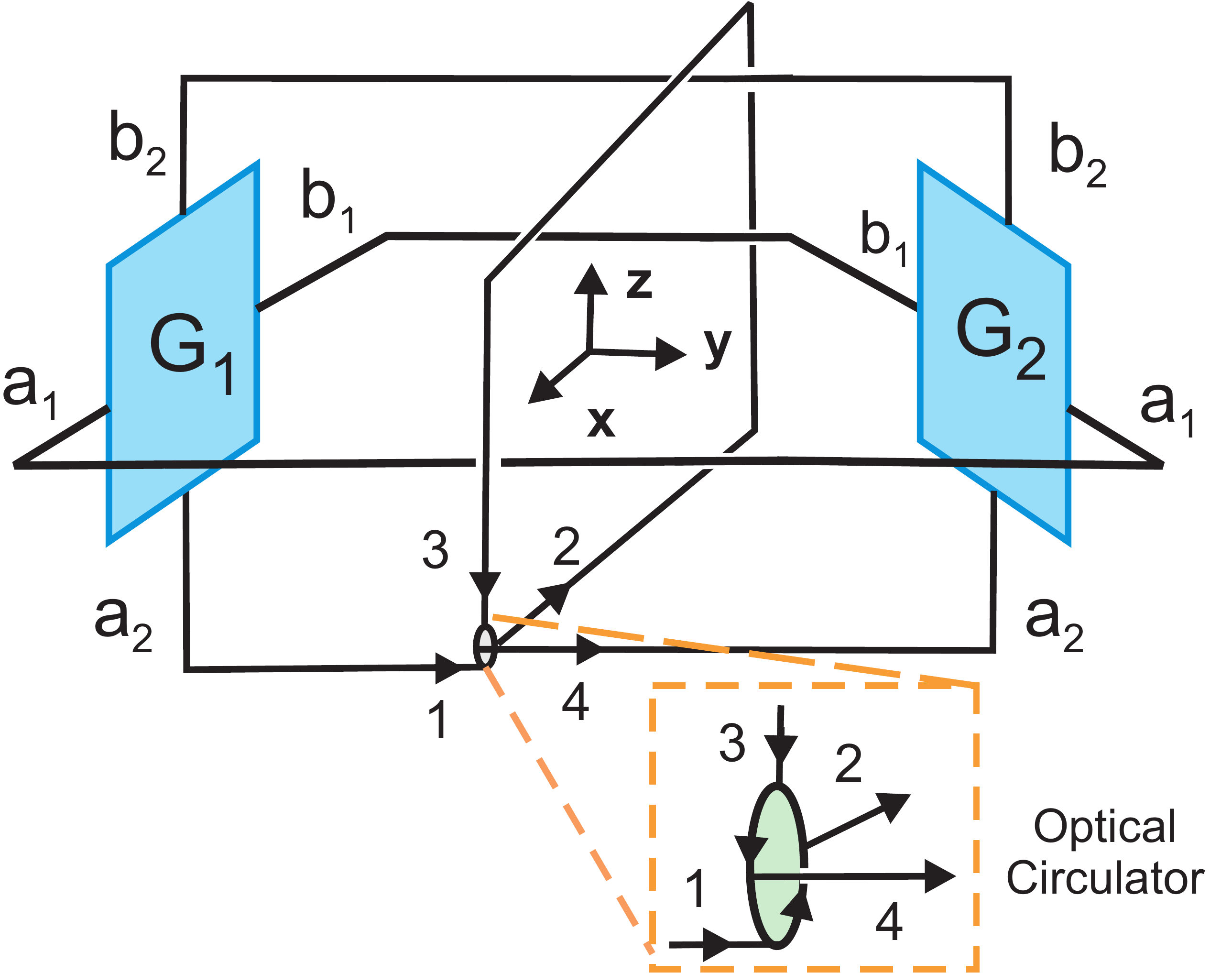}
\caption{The three-dimensional Grover-Sagnac interferometer: Measurement of rotations about a third axis can be added by altering the setup of the previous Figure using an optical circulator in one line to introduce a loop in the $y$-$z$ plane. Rotations of the system about the $y$-axis produce phase shifts proportional to the area of this loop. The circulator is shown expanded in the inset. Detectors are again understood to be coupled in via circulators after $G_2$ as in Fig. \ref{walkfig}.}\label{twistfig}
\end{figure}

To allow rotations about the $y$-axis to be measured as well, suppose now that an optical circulator is inserted into the $a_2$ line (Fig. \ref{twistfig}), allowing a loop in the $xz$-plane to be introduced into that line. Now, each passage from $G_1$ to $G_2$ exiting $a_1$ or $b_1$ comprises half a loop about the $z$-axis, and each passage from $G_1$ to $G_2$ leaving from $b_2$ comprises half a loop about $x$, as before, but each passage leaving $a_2$ introduces a half loop about $x$ \emph{and} a \emph{full} loop about $y$.

$\phi_0$, $\phi_1$, $\phi_2$ correspond, respectively, to closed loops exiting and reentering $G_1$ at ports $b_1$-$b_2$, $a_1$-$b_1$, and $a_2$-$b_2$, respectively, so the phase shifts measured by the coincidence rates in the previous section now correspond to phase shifts induced by rotations about the $x$, $y$, and $z$ axes through the relations:
\begin{eqnarray}\phi_0&=& {1\over 2} (\phi_x-\phi_z) \\ \phi_1&=& \phi_z \\
\phi_2&=&  \phi_x+\phi_y.   \end{eqnarray} In matrix form, \begin{equation}\left(\begin{array}{c} \phi_0\\ \phi_1\\ \phi_2 \end{array} \right) =\left(\begin{array}{ccc}  {1\over 2} & 0 &  -{1\over 2} \\ 0 & 0 & 1\\ 1 & 1 & 0 \end{array}\right)
\left(\begin{array}{c} \phi_x\\ \phi_y\\ \phi_z \end{array}\right) .\end{equation} Since the square matrix in the previous equation has nonzero determinant, measurement of $\phi_0$, $\phi_1$, $\phi_2$ via coincidence rates allows a unique solution for the rotation-induced shifts; namely:
\begin{eqnarray}\phi_x&=& 2\phi_0 +\phi_1 \\ \phi_y &=& -2\phi_0 -(\phi_1 + \phi_2) \\
\phi_z&=&  \phi_1.   \end{eqnarray}


Thus, by measuring $\phi_0,\; \phi_1,$ and $ \phi_2 $ via coincidence rates, as in the previous section, it is straightforward to reconstruct the rotation rates about all three axes simultaneously. The main drawback of this approach is that, due to the sign ambiguities mentioned in the previous section, in most cases only the rotation \emph{rate} (not the rotation \emph{direction}) can be determined.

It is useful to consider the resource requirements of the single proposed higher-dimensional interferometer, compared to three separate standard Sagnac interferometers. The Grover-based interferometer proposed here requires four detectors, whereas three Sagnacs would require three, meaning that only one additional detector would be needed. But consider that in applications where high resolution is essential, interferometers using quantum state input are useful, as mentioned in the introduction. A single Grover interferometer with one entangled-state input source would then certainly be less hardware-intensive than three standard interferometers, each with its own separate entangled-state input. In principle, one could try to use a single entangled source with a pair of beam splitters to direct the source output into the three interferometers. But with the three interferometers rotating independently along different axes, this would be very difficult to arrange in practice. In the Grover-based proposal, a single source shares the rotations of the single interferometer. So the resource savings due to having a single entangled source instead of three would more than compensate for the need for an extra detector. Thus, the proposal here may actually be feasible for ultra-high precision measurements, especially if the interferometers were fabricated on integrated chips. Such on-chip interferometers would remove potential stability and alignment problems.

\section{Conclusion}\label{conclusionsection}

In this paper we have shown that the use of linear-optical interferometers containing unbiased four-ports in place of beam splitters allows new effects to arise, such as continuous real-time tuning between HOM and anti-HOM effects, and the simultaneous measurements of multiple phase shifts, possibly induced by rotation rates, with reduced resources. These new possibilities arise from the increased dimensionality of the multiports and from their directionally-unbiased nature, allowing reflection backward toward the input direction. These interference effects are inherently quantum mechanical phenomena, requiring the use of two-photon states and coincidence counting. Whether additional interesting effects can be obtained in other interferometer topologies by similarly generalizing their vertices from beam splitters to directionally-unbiased multiports is an open avenue for future investigation.

As mentioned in the introduction, allowing the input optical states to be entangled allows improved phase resolution in many interferometers. The setups proposed here are naturally suited for such entanglement-based approaches; however, the effects of entanglement on the measurement resolution of these setups remain to be explored elsewhere.

\section*{Acknowledgements}
This research was supported by the Air Force Office of Scientific Research MURI award number FA9550-22-1-0312 and DOD/ARL grant number W911NF2020127.

\end{document}